\documentclass{aastex}          
\usepackage{spr-astr-addons}    

\begin{document}
%
\title{Dark energy coupled with dark matter in viscous fluid cosmology}

\shorttitle{Dark energy and dark matter}
\shortauthors{Brevik et al.}

\author{I. Brevik\altaffilmark{}}
\email{iver.h.brevik@ntnu.no}
 \affil{Department of Energy and Process Engineering, Norwegian University of Science and Technology, N-7491 Trondheim, Norway}

\and
\author{V. V. Obukhov\altaffilmark{}}
 \affil{Tomsk State Pedagogical University, 634061 Tomsk, Russia}
 \and
\author{A. V. Timoshkin\altaffilmark{}}
\affil{Tomsk State Pedagogical University, 634061 Tomsk, Russia}

\altaffiltext{1}{Affilation}

\begin{abstract}

We investigate  cosmological models with two interacting fluids:  dark energy and  dark matter in  flat Friedmann-Robertson-Walker universe. The interaction between dark energy and  dark matter is described in terms of the parameters present in the inhomogeneous equation of state when allowance is made for bulk viscosity,  for the Little Rip, the Pseudo Rip, and the bounce universes. We  obtain  analytic representation for characteristic properties in these cosmological models, in particular  the bulk viscosity $\zeta=\zeta(H,t)$ as function of Hubble parameter and time. We discuss the corrections of thermodynamical parameters in the equations of state due coupling between the viscous fluid and dark matter. Some common properties of these corrections are elucidated.
\end{abstract}


\section{ Introduction}

The appearance of new cosmological models is connected with the discovery of the accelerated expansion of the universe. Cosmic acceleration can be introduced via dark energy or via modification of gravity; cf.  \cite{nojiri11}. Recently, a general review of dark energy was given in \cite{bamba12}. Dark energy should have strong negative pressure and can be characterized by an equation of state parameter $w=p/\rho$, where $\rho$ is the dark energy density and $p$ is the dark pressure known to be negative. According to present observational data the value of $w$ is $w=-1.04^{+0.09}_{-0.10}$; cf. \cite{nakamura10}.
 There exist various cosmological scenarios for the evolution of the universe, called  the Big Rip; cf.  \cite{caldwell03,nojiri04},  the Little Rip; cf.  \cite{frampton11,brevik11,frampton12}, the Pseudo Rip; cf.  \cite{frampton12a}, the Quasi Rip; cf.  \cite{wei12}, and the bounce cosmology; cf.  \cite{novello08,bamba13,cai11}.
Cosmological models that treat dark energy and dark matter as imperfect fluids with unusual equation of state are considered in \cite{nojiri05,nojiri06},  where viscous fluids are just one particular case. Note that the case of a nonviscous (ideal) fluid is an idealized model useful in practice in many instances but in some situations incorrect.  It is necessary to take into account viscosity effects when  considering   turbulence effects; cf.  \cite{brevik12} or other realistic situations. From a hydrodynamic viewpoint the viscosity effect describes the deviation of the system from thermal equilibrium to the first order.  The influence of the bulk viscosity in the cosmic fluid plays an important role in the Big Rip phenomenon; cf.  \cite{brevik06,nojiri05a} or in  type II, III and IV Rip singularities; cf.  \cite{nojiri05a,brevik10}.
The Little Rip and the Pseudo Rip phenomena, and the bounce cosmology considered as an inhomogeneous nonviscous dark fluid coupled with dark matter, are considered in \cite{brevik13,brevik14}.
Cosmological models in which the modification of gravity is described  in terms of  a viscous fluid are explored in \cite{myrzakul14,myrzakulov14}.

The aim of this article is to obtain  analytical representations of the cosmological models induced by an inhomogeneous viscous fluid coupled with dark matter. This   realizes the Little Rip, the Pseudo Rip and the bounce scenarios via the thermodynamic parameter  $w(\rho)$ and the bulk viscosity in flat FRW space-time.

\section{Viscous cosmologies coupled with dark matter}

We consider a universe filled with two interacting fluid components: one dark energy component and one dark matter component in a spatially flat Friedmann-Robertson-Walker metric with scale factor $a$. The background equations; cf. \cite{nojiri11}, are
\begin{equation}
\left\{ \begin{array}{lll}
\dot{\rho}+3H(p+\rho)=-Q, \\
\dot{\rho}_m+3H(p_m+\rho_m)=Q, \\
\dot{H}=-\frac{k^2}{2}(p+\rho+p_m+\rho_m),
\end{array}
\right. \label{1}
\end{equation}
where $H=\dot{a}/a$ is the Hubble parameter and $k^2=8\pi G$,  $G$  being Newton's gravitational constant. Further, $p, \rho$ and $p_m,\rho_m$ are the pressure and the energy density of dark energy and dark matter correspondingly, and $Q$ is the interaction term between dark energy and dark matter. A dot denotes derivative with respect to cosmic time $t$.

Friedmann's equation for the Hubble parameter; cf.  \cite{nojiri11}, is
\begin{equation}
H^2=\frac{k^2}{3}(\rho+\rho_m). \label{2}
\end{equation}
We will now explore various cosmological models in which viscosity is coupled with dark matter.

\subsection{Little Rip model}
Little Rip cosmology is characterized by an energy density $\rho$ increasing with time but in an asymptotic sense, so that an infinite time is required to reach the  singularity. It corresponds to an equation-of-state parameter $w<-1$, but $w \rightarrow -1$ asymptotically. This a soft variant of  future singularity theory.

\bigskip
{\it First example.}

Let us consider the Little Rip model when the Hubble parameter has the form, cf.  \cite{frampton12},
\begin{equation}
H=H_0e^{\lambda t}, \quad H_0>0, \lambda>0, \label{3}
\end{equation}
where $H_0=H(0)$, $t=0$ denoting present time.

We will in the following assume dark matter to be dust matter, so that $p_m=0$. The gravitational equation of dark matter thus reduces to
\begin{equation}
\dot{\rho}_m+3H\rho_m=Q. \label{4}
\end{equation}
We take  the same coupling as in \cite{nojiri11},
\begin{equation}
Q=\delta H\rho_m, \label{5}
\end{equation}
where  $\delta$ is a positive nondimensional  constant.
The solution of Eq.~(\ref{4}) for dark matter is
\begin{equation}
\rho_m(t)=\tilde{\rho}_0\exp \left(\frac{\delta-3}{\lambda}H \right), \label{6}
\end{equation}
where $\tilde{\rho}_0$ is an integration constant. If $\delta$ is small ($<3$),
$\rho_m \rightarrow 0 $ when $t\rightarrow \infty$.

Let us consider the situation where the equation of state for the viscous fluid has the following inhomogeneous form; cf.  \cite{myrzakul14}
\begin{equation}
p=w(\rho)\rho-3H\zeta(H), \label{7}
\end{equation}
where $\zeta(H)$ is the bulk viscosity, dependent on the Hubble parameter. On thermodynamical grounds, one must have $\zeta(H) \geq 0$.

We now choose the thermodynamical (equation-of-state) parameter $w(\rho)$ to have the form; cf. \cite{myrzakul14}
\begin{equation}
w(\rho)=A_0\rho^{\alpha-1}-1, \label{8}
\end{equation}
where $A_0 \neq 0$ and $\alpha \geq 1$ are constants. Taking into account Eqs.~(\ref{2},\ref{3},\ref{7},\ref{8}) we the obtain the gravitational equation of motion for the dark energy
\begin{equation}
\frac{6H^3}{k^2}-\dot{\rho}_m+3H[A_0\rho^\alpha -3H\zeta(H)]=-\delta H\rho_m. \label{9}
\end{equation}
From this equation we obtain an expression for the bulk viscosity in the form
\begin{equation}
\zeta(H)=\frac{2}{3}\frac{\lambda}{k^2}+\frac{1}{3H}(\rho_m+A_0\rho^\alpha ).
\label{10}
\end{equation}
 The first term to the right in this expression is a constant; the second contains the contribution from the coupling.

In the limit $t \rightarrow \infty$, $H\rightarrow \infty$ exponentially, as does the density $\rho$ which is essentially proportional to $H^2$ according to Eq.~(\ref{2}). Thus the second term to the right in Eq.~(\ref{10}) goes to infinity in this limit.

Thus, we have formulated Little Rip cosmology theory in terms of the bulk viscosity concept, taking into account an interaction term $Q$ between dark energy and dark matter.

\bigskip
{\it Second example.}

Next, we will consider  another  Little Rip model where the Hubble parameter takes the form
\begin{equation}
H=H_0\exp(Ce^{\lambda t}), \label{11}
\end{equation}
where $H_0, C$ and $\lambda$ are positive constants. We see that in this case the parameter $H(t)$ increases much stronger with time than what was the case in Eq.~(\ref{3}).

 We take the interaction term $Q$ to have the same form (\ref{5}) as before, and consider the asymptotic for low values of $t$, implying $e^{\lambda t} \approx 1+\lambda t$.

Solving the gravitational equation of motion (\ref{4}) for dark matter, we find
\begin{equation}
\rho_m (t)=\tilde{\rho}_0\exp \left( \tilde{C}\,\frac{\delta-3}{\lambda}H\right). \label{12}
\end{equation}
Here $\tilde{\rho}_0$ is an integration constant and we have defined $\tilde{C}=C^{-1}e^C$.

We keep the same form (\ref{8}) for the dark energy thermodynamical parameter $w(\rho)$ in Eq.~(\ref{7}) as before. Then we obtain from the gravitational equation of motion (\ref{9}) for dark energy
\begin{equation}
\zeta(H,t)=\frac{2C\lambda}{3k^2}e^{\lambda t}+\frac{A_0\rho^\alpha}{3H}+\frac{\rho_m}{3H}\left[ \frac{\delta}{3}(1-e^C)+e^{C+\lambda t} \right]. \label{13}
\end{equation}
This concludes our second variant of a bulk viscosity  Little Rip model.

\subsection{Pseudo Rip model}

In Pseudo Rip cosmology the Hubble parameter tends to the cosmological {\it constant} in the remote future. That means, the universe approaches a de Sitter space. We will now analyze this model in analogy with the above Little Rip model.

\bigskip
{\it First example. }

Let us suppose that the Hubble parameter has the form; cf. \cite{frampton12}
\begin{equation}
H=H_0-H_1\exp(-\tilde{\lambda}t), \label{14}
\end{equation}
where $H_0, H_1$ and $\tilde{\lambda}$ are positive constants, $H_0>H_1, t>0$.
 For low values of $t$ we have $H\rightarrow H_0-H_1$, and in the late-time universe $H\rightarrow H_0$. The Hubble parameter approaches a constant.

Taking the interaction between dark energy and dark matter to have the form (\ref{5}), we find as solution of the gravitational equation (\ref{4})
\begin{equation}
\rho_m(t)=\tilde{\rho}_0\exp \left[ (\delta-3)\left(H_0t-\frac{H-H_0}{\tilde{\lambda}}\right) \right], \label{15}
\end{equation}
where $\tilde{\rho}_0$ is an integration constant.

If the thermodynamical parameter $w(\rho)$ has the form (\ref{8}) we obtain from Eq.~(\ref{9}) the following expression for the bulk viscosity:
\begin{equation}
\zeta(H,t)=\frac{2\tilde{\lambda}}{3k^2}\left( \frac{H_0}{H}-1\right)+\frac{1}{3H}(\rho_m+A_0\rho^\alpha ). \label{16}
\end{equation}
The bulk viscosity contains the coupling, as before.

\bigskip
{\it Second example.}

This example considers a cosmological model with asymptotic de Sitter evolution.

The Hubble parameter is taken in the form; cf. \cite{frampton12}
\begin{equation}
H=\frac{x_f}{\sqrt 3 }\left[ 1-\left(1-\frac{x_0}{x_f}\right) \exp \left( -\frac{\sqrt{3}At}{2x_f}\right)\right], \label{17}
\end{equation}
where $x_0=\sqrt{\rho_0}$ corresponds to the present energy density, $x_f$ is a finite quantity, and $A$ is a positive (dimensional) constant. When $t\rightarrow \infty$ the Hubble parameter $H\rightarrow x_f/\sqrt 3$, and the expression (\ref{17}) tends asymptotically to the de Sitter solution. When $t\rightarrow 0$, we obtain $H\rightarrow  x_0/\sqrt 3$.

For the energy density of dark matter we find
\begin{equation}
\rho_m(t)=\tilde{\rho}_0\exp \left\{ (\delta-3)\frac{x_f}{\sqrt 3}\left[ t+\frac{2x_f}{\sqrt{3}A}\left[1-\frac{\sqrt 3}{x_f}H\right)\right]\right\}, \label{18}
\end{equation}
where $\tilde{\rho}_0$ is an integration constant.

If we repeat the above procedure again, we obtain the following expression for the bulk viscosity in this model:
\begin{equation}
\zeta(H,t)=\frac{A}{3k^2H}\left( 1-\frac{\sqrt 3}{x_f}H\right)+\frac{1}{3H}(\rho_m+A_0\rho^\alpha). \label{19}
\end{equation}

Thus we see that, unlike the case where only a dark energy is involved, the interaction between dark energy and dark matter in the gravitational equations leads to corrections in the bulk viscosity.

\subsection{Bounce cosmology}

There have appeared  several papers discussing  the possibility of  a matter bounce cosmology in the early universe; cf. \cite{novello08,bamba13,cai11}. In this model the universe goes from an era of accelerated collapse to the expanding era without displaying a singularity. This is essentially a cyclic universe model. After the bounce the universe soon enters a matter-dominated expansion phase. In the following we will study cosmological models of basically this type.

\bigskip

{\it Exponential model.}

Let us consider a bounce cosmological model for which the scale factor $a$ is given in exponential form; cf. \cite{myrzakulov14}:
\begin{equation}
a(t)=a_0\exp\left[ \beta (t-t_0)^{2n}\right], \label{20}
\end{equation}
where $\alpha_0, \beta$ are positive (dimensional) constants, $n \in N$, and $t_0$ the fixed bounce time. The Hubble parameter behaves as
\begin{equation}
H(t)=2n\beta (t-t_0)^{2n-1}. \label{21}
\end{equation}
When $t<t_0$, the scale factor decreases and there occurs a contraction of the universe. At $t=t_0$ when $a_0=a(t_0)$, the bounce takes place. After the bounce, when $t>t_0$, the scale factor increases and the universe expands.

We choose the coupling with dark matter to have the same form (\ref{5}) as before. Then we find from Eq.~(\ref{4}) the energy density for dark matter
\begin{equation}
\rho_m(t)=\tilde{\rho}_0\exp \left[ \frac{(\delta-3)}{2n}H(t)(t-t_0)\right], \label{22}
\end{equation}
where $\tilde{\rho}_0$ is an integration constant.

If the equation of state for the fluid has the form (\ref{7}) and the thermodynamical parameter $w(\rho)$ is as in Eq.~(\ref{8}), we obtain the following expression for the bulk viscosity for the bounce solution:
\begin{equation}
\zeta(H,t)=\frac{2(2n-1)}{3k^2(t-t_0)}+\frac{1}{3H}(A_0\rho^\alpha+\rho_m). \label{23}
\end{equation}
This solution also contains the coupling correction.

\bigskip
{\it Power-law model.}

Now we will analyze the bounce cosmological scenario when the scale factor has the following form; cf. \cite{myrzakulov14}:
\begin{equation}
a(t)=a_0+\beta(t-t_0)^{2n}, \label{24}
\end{equation}
where as before $\alpha_0, \beta$ are dimensional constants, $n\in N$, and $t=t_0$ is the fixed bounce time.

The Hubble parameter becomes
\begin{equation}
H(t)=\frac{2n\beta (t-t_0)^{2n-1}}{a_0+\beta(t-t_0)^{2n}}. \label{25}
\end{equation}
If the assumptions (\ref{5}), (\ref{7}), (\ref{8}) are fulfilled, the solution of the energy conservation law (\ref{4}) for dark matter becomes
\begin{equation}
\rho_m(t)=\tilde{\rho}_0[a(t)]^{\delta-3}. \label{26}
\end{equation}
Then we  obtain from the gravitational equation of motion (\ref{9}):
\begin{equation}
\zeta(H,t)=\frac{2}{3k^2}\left( \frac{2n-1}{t-t_0}-H\right)+\frac{1}{3H}(\rho_m+A_0\rho^\alpha). \label{27}
\end{equation}
This is the bulk viscosity, corresponding to the power-law model.

\section{Remarks}

It is possible to rewrite the expressions for the bulk viscosity in Section 2 in the following general form:
\begin{equation}
\zeta(H,t)=\zeta_0(H,t)+\frac{1}{3H}\left\{ [w(\rho)+1]\rho+\rho_m\right\}. \label{28}
\end{equation}
Here the first term is connected with the specific model; the second term contains the coupling correction and depends also on the choice of thermodynamical parameter.

The Little Rip, Pseudo Rip, and bounce cosmological models may be related to each other via the bulk viscosity concept. As an example, we may consider the following transactions:

If we choose in Eq.~(\ref{10}) the parameter $\lambda$ to be $\lambda=(2n-1)/(t-t_0)$ and choose in Eq.~(\ref{21}) $\beta=\frac{H_0}{2n}\left[ \frac{\exp[t/(t-t_0)]}{t-t_0}\right]^{2n-1}$, we obtain an expression which describes the possible transition from the Little Rip model to the bounce model. Analogously, by choosing $\frac{\tilde{\lambda}}{(H_0/H_1)\exp(\tilde{\lambda}t)-1}=\frac{2n-1}{t-t_0}$ in Eq.~(\ref{16}) and $\beta=\frac{H_0-H_1\exp(-\tilde{\lambda}t)}{2n(t-t_0)^{2n-1}}$ in Eq.~(\ref{21}), we find the possible transition from the Pseudo Rip to the bounce model.

\section{Conclusion}
In this paper we have studied examples of Little Rip, Pseudo Rip, and bounce cosmology, described in flat Friedmann-Robertson-Walker spacetime when the cosmic fluid is  viscous and is coupled with dark matter. We have found corrections in the thermodynamical parameter, and to the bulk viscosity, in the equation of state for the dark energy.

It should be mentioned that there are common properties of the corrections present in these models. Thus $\Delta \zeta_m(H,t)=\rho_m/(3H)$ represents the correction to the bulk viscosity from the coupling. Further, $\Delta \zeta_\rho(H,t)=(A_0/3H)\rho^\alpha$ is related to the choice of thermodynamical parameter $w(\rho)$ in the equation of state.

From a general perspective  this paper may be considered as inhomogeneous cosmic fluid theory with changeable thermodynamical parameter $w$ coupled to dark matter \cite{elizalde14}, applied to the Little Rip, the Pseudo Rip, and the bounce phenomena.

It is known, cf. \cite{nojiri11}, that a viscous fluid may be understood also as a modified gravity model,
for instance of the $F(R)$ type. It is moreover known that $F(R)$ gravity may provide a unification of early-time inflation with a special version of dark energy, as was proposed by  \cite{nojiri03}, or with the nonlinear model proposed by \cite{nojiri08}.     Having that in mind, we expect that there is a natural possibility to unify these epochs with inflation in an extended viscous model. This will be considered elsewhere.

\section* {Acknowledgments}

This work was supported by a grant from the Russian Ministry of Education and Science, project TSPU-139 (A.V.T. and V.V.O.).

%



\bibliographystyle{spr-mp-nameyear-cnd}

 \bibliography{<bib data>}                

%

\end{document}